\documentclass[conference]{IEEEtran}
\makeatletter
\def\ps@headings{%
    \def\@oddhead{\mbox{}\scriptsize\rightmark \hfil \thepage}%
    \def\@evenhead{\scriptsize\thepage \hfil \leftmark\mbox{}}%
    \def\@oddfoot{}%
    \def\@evenfoot{}}
\makeatother
\usepackage[top=0.76in, bottom=1in, left=0.63in, right=0.66in]{geometry}
\usepackage{paralist}
\usepackage{pgf}
\usepackage{epstopdf}
\usepackage{etex}
\usepackage{tikz}
\usetikzlibrary{arrows,automata}
\usepackage{algcompatible}
\usepackage{algorithm}
\usepackage{soul}
\usepackage{amsmath}

\usepackage{amssymb}
\usepackage{tikz-qtree}
\usepackage{array}
\usepackage[american]{babel}
\usepackage{graphicx}
\usepackage{subfigure}
\usepackage{amssymb}
\usepackage{enumitem}
\usepackage{algorithm}
\usepackage{url}
\usepackage{xcolor,listings}
\usepackage{pgfplots}
\usepackage{verbatimbox}
\usepackage{multirow}
\usepackage[section]{placeins}
\usepackage{tikz}

\pgfplotsset{every axis legend/.style={
        cells={anchor=center},
        inner xsep=3pt,
        inner ysep=2pt,
        nodes={inner sep=2pt,text depth=0.15em},
        anchor=south east,
        shape=rectangle, 
        fill=white, 
        draw=white, 
        at={(rel axis cs:0.50,0.65)} 
}}

\usepackage{amsthm}

\usepackage{algorithm}
\usepackage[noend]{algpseudocode}
 
\usepackage{forest}

\usetikzlibrary{matrix,calc,positioning}
\usetikzlibrary{arrows.meta, shapes.geometric, calc, shadows}

\colorlet{mygreen}{green!75!black}
\colorlet{col1in}{red!30}
\colorlet{col1out}{red!40}
\colorlet{col2in}{mygreen!40}
\colorlet{col2out}{mygreen!50}
\colorlet{col3in}{blue!30}
\colorlet{col3out}{blue!40}
\colorlet{col4in}{mygreen!20}
\colorlet{col4out}{mygreen!30}
\colorlet{col5in}{blue!10}
\colorlet{col5out}{blue!20}
\colorlet{col6in}{blue!20}
\colorlet{col6out}{blue!30}
\colorlet{col7out}{orange}
\colorlet{col7in}{orange!50}
\colorlet{col8out}{orange!40}
\colorlet{col8in}{orange!20}
\colorlet{linecol}{blue!60}

\usepackage{listings}
\usepackage{xcolor}
\usepackage{float}
\usepgfplotslibrary{groupplots}
\colorlet{punct}{red!60!black}
\definecolor{background}{HTML}{EEEEEE}
\definecolor{delim}{RGB}{20,105,176}
\colorlet{numb}{magenta!60!black}

\lstdefinelanguage{json}{
    basicstyle=\normalfont\ttfamily,
    numbers=left,
    numberstyle=\scriptsize,
    stepnumber=1,
    numbersep=8pt,
    showstringspaces=false,
    breaklines=true,
    frame=lines,
    backgroundcolor=\color{background},
    literate=
     *{0}{{{\color{numb}0}}}{1}
      {1}{{{\color{numb}1}}}{1}
      {2}{{{\color{numb}2}}}{1}
      {3}{{{\color{numb}3}}}{1}
      {4}{{{\color{numb}4}}}{1}
      {5}{{{\color{numb}5}}}{1}
      {6}{{{\color{numb}6}}}{1}
      {7}{{{\color{numb}7}}}{1}
      {8}{{{\color{numb}8}}}{1}
      {9}{{{\color{numb}9}}}{1}
      {:}{{{\color{punct}{:}}}}{1}
      {,}{{{\color{punct}{,}}}}{1}
      {\{}{{{\color{delim}{\{}}}}{1}
      {\}}{{{\color{delim}{\}}}}}{1}
      {[}{{{\color{delim}{[}}}}{1}
      {]}{{{\color{delim}{]}}}}{1},
}

\ifCLASSINFOpdf
\else
\fi

\begin{document}
\newtheorem{mydef}{Definition}

%
\title{SUPC: SDN enabled Universal Policy Checking in Cloud Network}

\author{
    \IEEEauthorblockN{Ankur Chowdhary, Adel Alshamrani, and Dijiang Huang}
    \IEEEauthorblockA{Arizona State University
        \\\{achaud16, aalsham4, dijiang\}@asu.edu}
}


%


\maketitle

\begin{abstract}
    Multi-tenant cloud networks have various security and monitoring service functions (SFs) that constitute a service function chain (SFC) between two endpoints. SF rule ordering overlaps and policy conflicts can cause increased latency, service disruption and security breaches in cloud networks. Software Defined Network (SDN) based Network Function Virtualization (NFV) has emerged as a solution that allows dynamic SFC composition and traffic steering in a cloud network. We propose an SDN enabled Universal Policy Checking (SUPC) framework, to provide 1) Flow Composition and Ordering by translating various SF rules into the OpenFlow format. This ensures elimination of redundant rules and policy compliance in SFC. 2) Flow conflict analysis to identify conflicts in header space and actions between various SF rules. Our results show a significant reduction in SF rules on composition. Additionally, our conflict checking mechanism was able to identify several rule conflicts that pose security, efficiency, and service availability issues in the cloud network.
\end{abstract}

\IEEEpeerreviewmaketitle

\section{Introduction}

\noindent Service Functions (SFs) comprise a class of middleboxes, such as  Firewall, Intrusion Detection System (IDS) and Deep Packet Inspection (DPI), that examine and modify the traffic and flows in a sophisticated fashion. The SFs at the network level is known as Network Functions (NFs)~\cite{gember2014opennf}. Network Function Virtualization (NFV) proposes replacement of hardware middleboxes with flexible and programmable software middleboxes, which is referred to as Virtual Network Functions (VNFs)~\cite{ghaznavi2016service}. 

\noindent  Lack of a common protocol standard among middleboxes makes it hard to debug the configuration errors. In case of any middlebox failure, network administrators rely on ad-hoc rules to resolve the issue, which lacks a comprehensive view of the overall network and are error prone~\cite{durante2017model}. 
\noindent Current Service Function Chaining (SFC) solutions that consider VNF policy ordering and policy compliance~\cite{joseph2008policy, gember2014opennf} do not analyze the policies at the granularity of the packet header. The policy specifications of SF and actual implementation on the network are often quite different, and without a common standard to interpret various SFs, the ordering mechanism is hard to be verified.  Network-wide policy enforcement solution for middleboxes discussed in FlowTags~\cite{fayazbakhsh2013flowtags} may not scale well on a large cloud network since tracking of tags across various VNFs and backtracking in case of SF failure can be quite difficult. The policy composition mechanisms such as PGA~\cite{prakash2015pga}, implement SFs multiple times in order to achieve the desired security objectives, however, there is a significant overlap in the packet header and actions of SFs, which slows down the SF performance. The solutions that focus exclusively on security enforcement in SFC~\cite{sendi2016efficient, wang2016sics} do not consider the overlap between the packet header and the associated actions in SFC, which can cause policy level conflicts. 

\noindent  SDN has the capability of dynamically managing VNF connections and data plane flows as discussed by Trajkovska \textit{et al}.~\cite{trajkovska2017sdn}. SDN utilizes OpenFlow~\cite{mckeown2008openflow} (programmable network protocol) for interacting with forwarding plane of network devices~\cite{mckeown2008openflow}. In the SUPC framework, we reduce all SF configuration to OpenFlow rules in order to have a global view of the network. This SDN controller OpenDaylight in our framework provides end-to-end visibility to help resolve misconfiguration related failures.  SUPC framework utilizes an OpenFlow format to identify conflicting scenarios across different SFs. The presence of common rule format also helps in automatic verification of security policies across various SFs. The key contributions of this research work are as follows:
\begin{itemize}
    \item SUPC utilizes packet header fields and traffic steering of SFs and to composes a set of OpenFlow rules with no duplicates. We incorporate correct service ordering by assigning priority to flow rules derived from SFs with higher precedence. We were able to achieve a significant reduction in the number of rules to be processed and the flow composition time in our experimental analysis. 
    \item SUPC identifies four major type of rule conflicts based on important network and security properties. We identified 100s of conflicts in our SDN cloud dataset with 25k OpenFlow rules, which can cause security conflicts and network service failures.
\end{itemize}

\section{Related Work}
\noindent  \textbf{Security in SFC} has been modeled in SICS~\cite{wang2016sics}. The authors consider rule composition, header space mapping, order and priority of various requirements specified as part of SFC. The algorithm is however based on simple predicate logic, which doesn't allow deductive reasoning to make some inference about rule conflicts as in case of our work. Network security defense pattern (NSDP) based on multiple objectives such as security, minimal resource wastage, fault tolerance has been proposed by Sendi \textit{et al.}~\cite{sendi2016efficient}. Authors use mechanisms such as decomposition, composition, location and zone awareness for service chain composition. Our work uses a constrained set of OpenFlow rule representation which will be more cost efficient in terms of SFC latency compared to this solution.  

\noindent  \textbf{Policy Aware} SFC has been discussed by Joseph \textit{et al.}~\cite{joseph2008policy}. Authors separate policy from reachability in SFC in order to ensure correctness and flexibility. FlowTags~\cite{fayazbakhsh2013flowtags} extends \textit{SDN} architecture for adding tags to outgoing packets. This provides a necessary context for policy enforcement. These works do not, however, consider possible overlaps between network policies based on packet header match. Works that focus on policy safety and efficiency of SFC like SDN based virtual firewall discussed by Brew~\cite{7976378} consider issues like semantic consistency, and scalability but their application is limited to the firewall SF. 



\section{Background}
\noindent  \textbf{SDN based SFC:} We consider the multi-tenant cloud network as a use case to highlight some of these issues in detail.  In this example, we consider two Compute nodes with four SFs namely \textit{Classifier (C), IDS, VPN and Load Balancer (LB)}. We discuss a few ordering and placement strategies and highlight the shortcomings below. Our SFC has the following requirements:
\begin{enumerate}
    \item \textit{Traffic coming into the network should be classified into different categories based on source IP address using Classifier SF.}
    \item \textit{Any traffic not part of data network security domain should be processed via Intrusion Detection System.} 
    \item \textit{Data network traffic and SDN controller traffic should go through Load balancing SF.}
    \item \textit{Control plane traffic from SDN controller should be encrypted using public key encryption scheme.}
\end{enumerate}

\begin{figure}[ht!]
    \centering
    \includegraphics[width=0.45\textwidth]{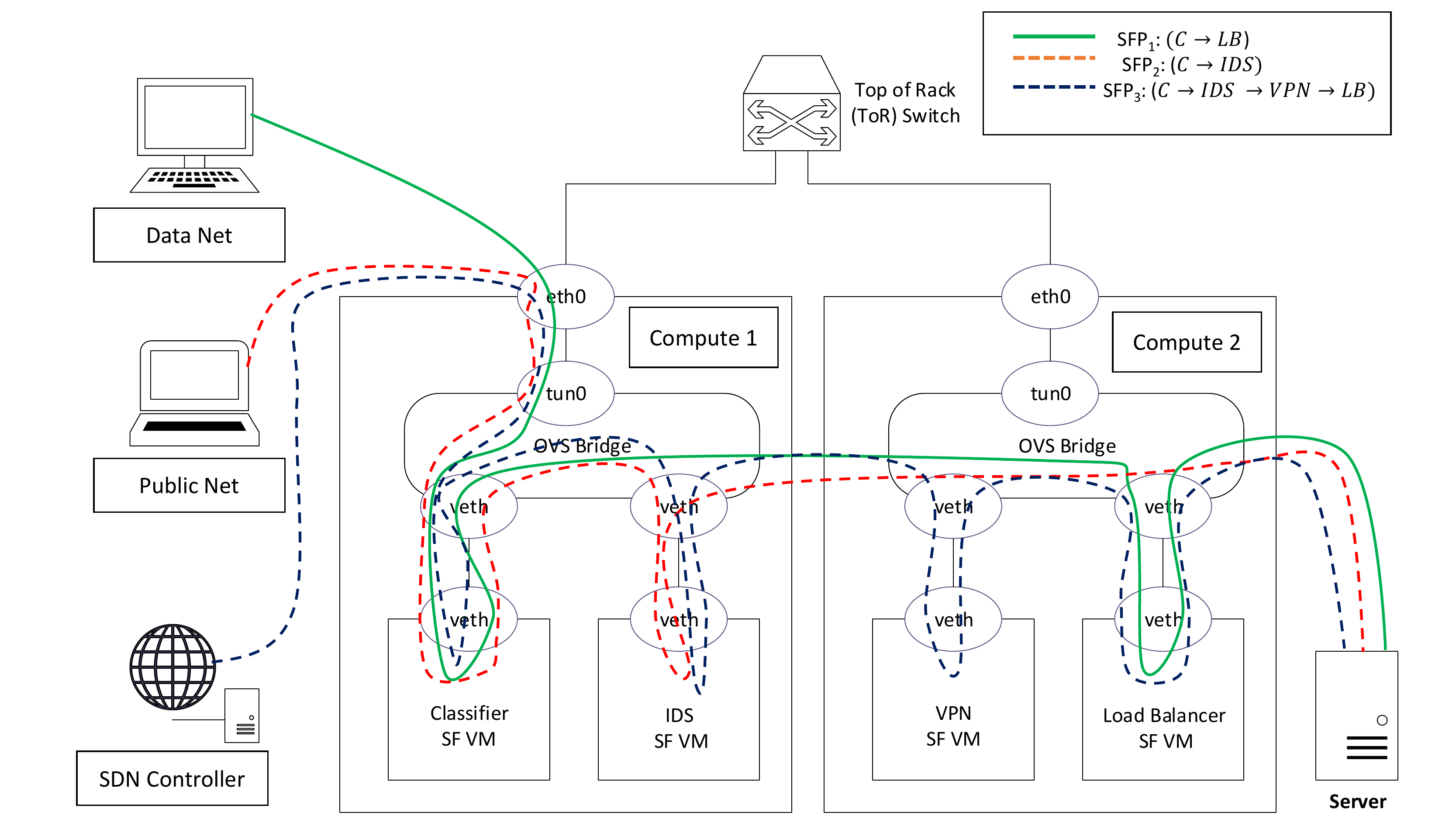}
    \caption{Service Function Chaining Example in Cloud Network}
    \label{fig:sfc02}
\end{figure}
\noindent  The nodes in an SFC architecture can be SFs or \textit{Service Function Forwarders (SFFs)}. An SFF is responsible for forwarding traffic packets or frames received from a particular network segment to associated SFs using the information encapsulated in the packet. In the Figure~\ref{fig:sfc02} the Open vSwitch (OVS) bridge acts as SFF. 

\noindent The \textit{Service Function Path (SFP)} is the actual path traversed by the packet/frame from source to destination in SFC after application of granular policies and operational constraints in SFC. For instance in the Figure~\ref{fig:sfc02}, there are three SFPs - $SFP_1, SFP_2, SFP_3$ corresponding to Data Net, Public Net and SDN controller traffic.  

\noindent \textbf{Strategy 1} \textbf{Order:} $C \rightarrow VPN \rightarrow IDS \rightarrow LB$. \\
\textit{Issue:} SDN controller traffic needs to go through both VPN and IDS as per policy, placing VPN first (incorrect order) violates security objective. Thus, IDS should precede VPN, since VPN encrypts the traffic and IDS can operate only on the raw traffic.

\noindent \textbf{Strategy 2} \textbf{Order:} $C \rightarrow LB \rightarrow IDS \rightarrow VPN$. \\
\textit{Issue:} The traffic from SDN controller and data network has to go through classifier and load balancer. Malicious traffic could have been filtered out using IDS policy resulting in less impact on QoS offered by the load balancer. The incorrect placement leads to an efficiency issue. In order to preserve both security and efficiency constraints, we design better placement and ordering as shown in Fig.~\ref{fig:sfc02} - \textit{Strategy 3}. 

\noindent \textbf{Strategy 3} \textbf{Order:} $C \rightarrow IDS \rightarrow VPN \rightarrow LB$. \\
\textit{Efficient Placement} is obtained in this strategy since unwanted traffic is filtered at IDS and load balancer has to deal with only legitimate traffic from \textit{Data-Net} and \textit{SDN Controller}. \\
\textit{Correct Ordering} is obtained for SDN controller traffic. The traffic is passed in raw (un-encrypted) format through IDS, and later through VPN thus IDS has complete visibility.

\begin{figure}[ht!]
    \centering
    \includegraphics[trim=0 40 0 50, width=0.45\textwidth]{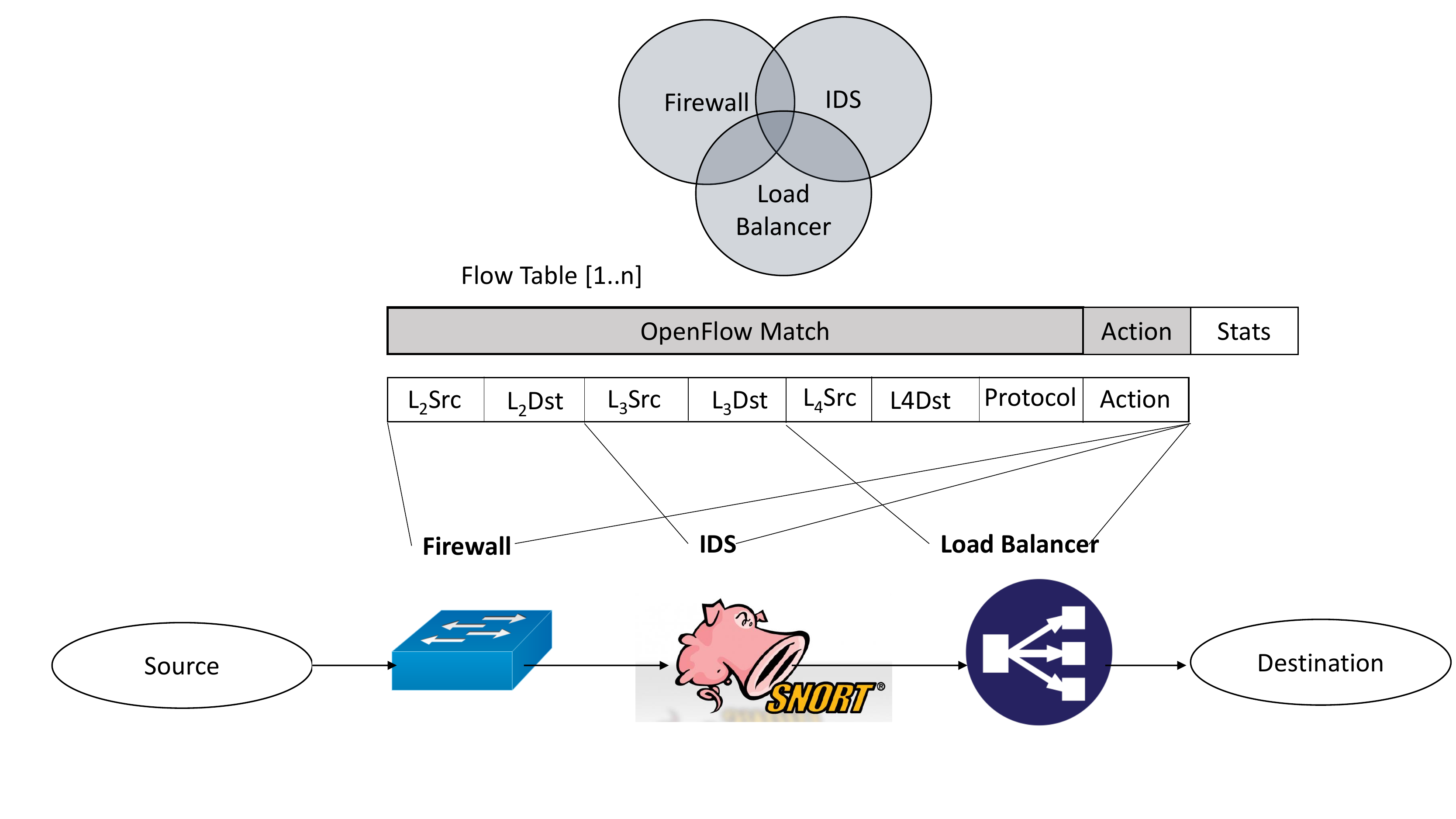}
    \caption{Packet Header space Overlap}
    \label{fig:sfc08}
\end{figure}
\noindent  Creation of SFC can lead to several options. We observed that current deployments of SFs in current works lack following desirable properties, which leads to the presence of redundant and conflicting rules in SFC: 

\noindent \textbf{(i) SF Rule Ordering and Composition:} Figure~\ref{fig:sfc08} shows that the packet header space of Firewall, IDS, and Load Balancer in SFC overlap with each other. Inefficient placement of SFs can incur communication cost on the network. For instance, if IDS was placed before Firewall, IDS would have to analyze traffic which may be dropped by Firewall in the current arrangement. 
\noindent \textbf{(ii) SF Conflict Analysis:} We identified several conflicting scenarios in security and traffic processing policies in SFC after the composition of SF rules into OpenFlow format due to partial or full overlap in packet match and action fields of SF rules. These conflicts can cause security policy violations due to conflicting actions and service disruption because some symmetric traffic flows require configuration changes for incoming and outgoing traffic. 

\section{SFC Composition and Conflict Checking}
The possible overlap in packet header provides scope for policy composition and traffic steering in an efficient fashion. We automate flow rule match (layer 2-4 headers) and action set composition to OpenFlow rules at each middlebox in SFC.
\subsection{Flow Composition}
\noindent  To illustrate the flow composition problem we take example of traffic manipulation and access control policies of some SFs and their corresponding translation into OpenFlow rules. In OpenFlow specification~\cite{mckeown2008openflow}, each flow rule is a tuple of three sets, i.e. - \textit{match}, \textit{action (A)} and \textit{priority (P)}. We define each flow rule $R_i = \{match_i,A_i,P_i\}$. The match field for each rule $R_i$ consists of several sub-fields such as source MAC address $l2s_i$, destination MAC address $l2d_i$ , source IP address $l3s_i$, destination IP address $l3d_i$, source port $l4s_i$, destination port $l4d_i$, protocol $\rho_i$. We consider seven sub fields for our conflict detection model. Thus, $match(R_i)$ = $\{l2s_i, l2d_i, l3s_i, l3d_i, l4s_i, l4d_i, \rho_i\}$.

\begin{figure}[ht!]
    \centering
    \includegraphics[trim=0 40 0 50, width=0.45\textwidth]{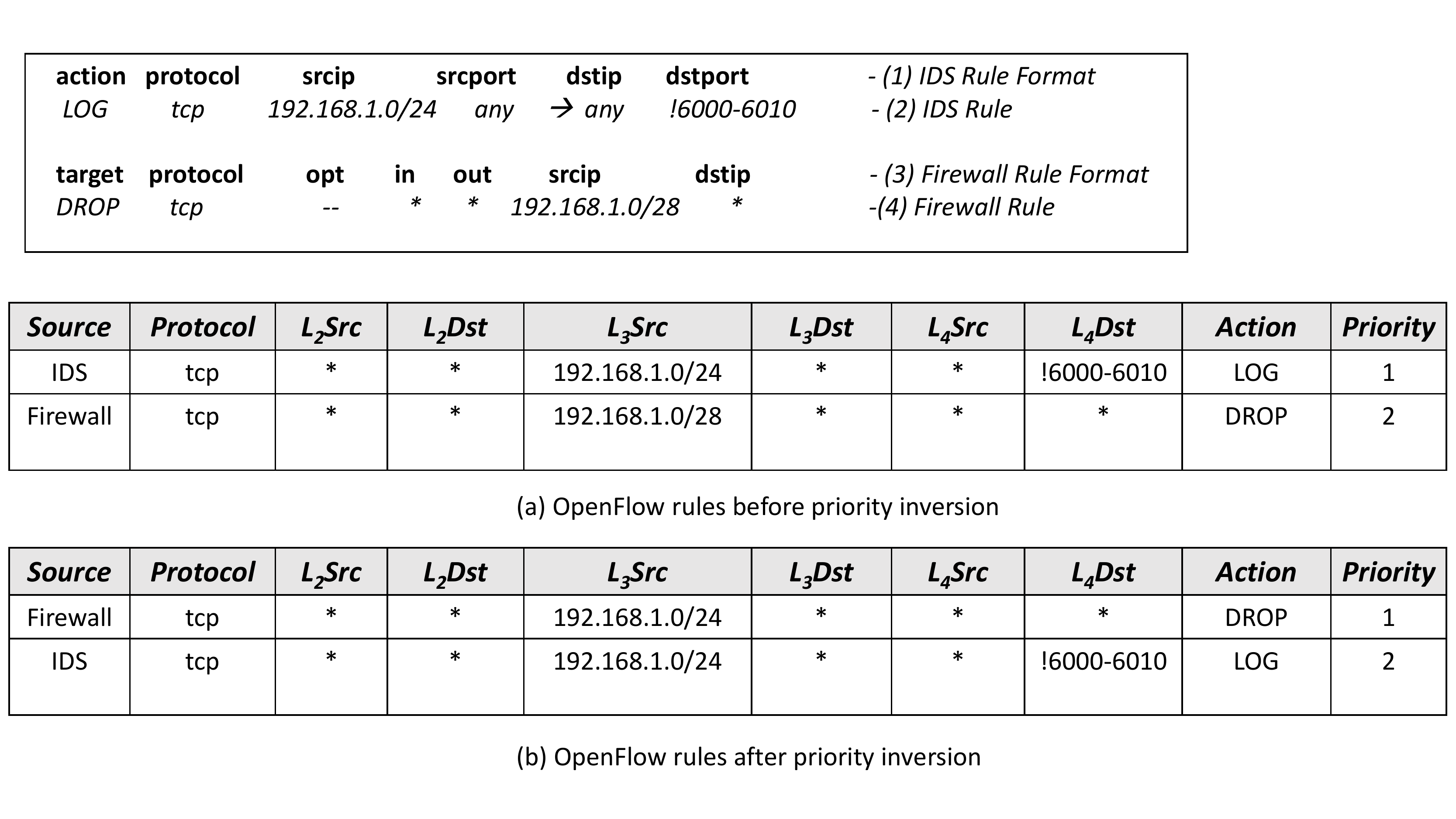}
    \caption{Flow Composition Example}
    \label{fig:sfc05}
\end{figure}

\noindent Consider a  Snort IDS rule - line 2 and Firewall rule line 4 in Figure~\ref{fig:sfc05}. Figure~\ref{fig:sfc05}(a) shows the OpenFlow rules composed from security policy rules. The IDS rule and the Firewall rule can be translated to OpenFlow rule using one to one mapping of header fields, e.g., $srcip_{IDS}$  and  $dstip_{IDS}$ are added as \textit{L3 Src} and \textit{L3 Dst} in the flow table of OpenFlow switch. Similarly, srcport and dstport of IDS rules are added as L4 Src and Dst. 
The module also checks for the ordering of SFs based on the source of flow rules. As can be seen from the Figure, the IDS has header space overlap because the layer 3 source and destination of IDS and Firewall are overlapping, i.e. \{$L3_{Firewall} \subset L3_{IDS}$\}. The IDS will have to handle additional traffic, which is matching the header space of the firewall rule. Ideally, the firewall rule should be applied before the IDS, so that traffic corresponding to $L3_{Firewall}$ is dropped and only the traffic from the set \{$match_{Firewall} \setminus match_{IDS}$\} is processed by IDS. The flow composition, in our SFC implementation, inverts the priority of the rules,  Figure~\ref{fig:sfc05}(b).

\begin{algorithm}[ht!]
    \caption{Flow Rule Composition}\label{alg01}
      \begin{small}
    \begin{algorithmic}[1]
        \Procedure{Flow Rule Compilation}{\texttt{R, S}}
        \State $R \gets$ HashSet
        \State $S \gets$ service function rules
        \State $s \in$ S \Comment{All Netfilter, proxy, IDS rules in S}
        \State $i \in \{1,n\}$ 
        \State $P_{FW} \gets$ Priority Set

        \For{$s_i \in$ S}
            \State $R_i$.match().prot() $\gets s_i$.prot()
            \State $R_i$.match().ipSrc() $\gets s_i$.ipSrc()   
            \State $R_i$.match().ipDst() $\gets s_i$.ipDst() 
             \State $R_i$.match().sPort() $\gets s_i$.sPort()  
            \State $R_i$.match().dPort() $\gets s_i$.dPort() 
            \State $R_i$.action() $\gets$ $s_i$.action() 
        \If{$s_i \in$ S.Netfilter()}
             \State $R_i$.match().hwSrc() $\gets s_i$.hwSrc()  
            \State $R_i$.match().hwDst() $\gets s_i$.hwDst()
            \State $R_i$.prior() $\gets$ rand(1,65535)
            \State $P_{FW}.add(R_i.prior())$
         \ElsIf{$s_i \in$ S.IDS()}
            \State $R_i$.prior() $\gets$ rand(max($P_{FW}$, 65535))
        \EndIf    
         
        \EndFor
        \EndProcedure
        
    \end{algorithmic}
\end{small}

\end{algorithm}
\noindent  In the algorithm~\ref{alg01}, we current rule list from various network functions is represented by $S$. In the lines 7-13, we perform one-one mapping of the protocol defined in SF rule, layer 3 sources and destination address, layer 4 sources, and destinations addresses. In the 14 we check the source of SF, i.e., firewall or IDS. If the source is a firewall, we add layer 2 sources and destination address into flow rule we are composing - lines 15-16. We assign a random priority - line 17 from 1-65535 (upper limit of OpenFlow rule priority). If the source is IDS - line 19, we assign a priority value to flow rule greater than firewall rule set, so when OpenFlow rules are processed, firewall rules gets precedence over IDS rule as discussed in example  Figure~\ref{fig:sfc05}.


\subsection{Conflict Checking}

\begin{figure}[ht!]
    \centering
    \includegraphics[trim=0 30 0 50, width=0.45\textwidth]{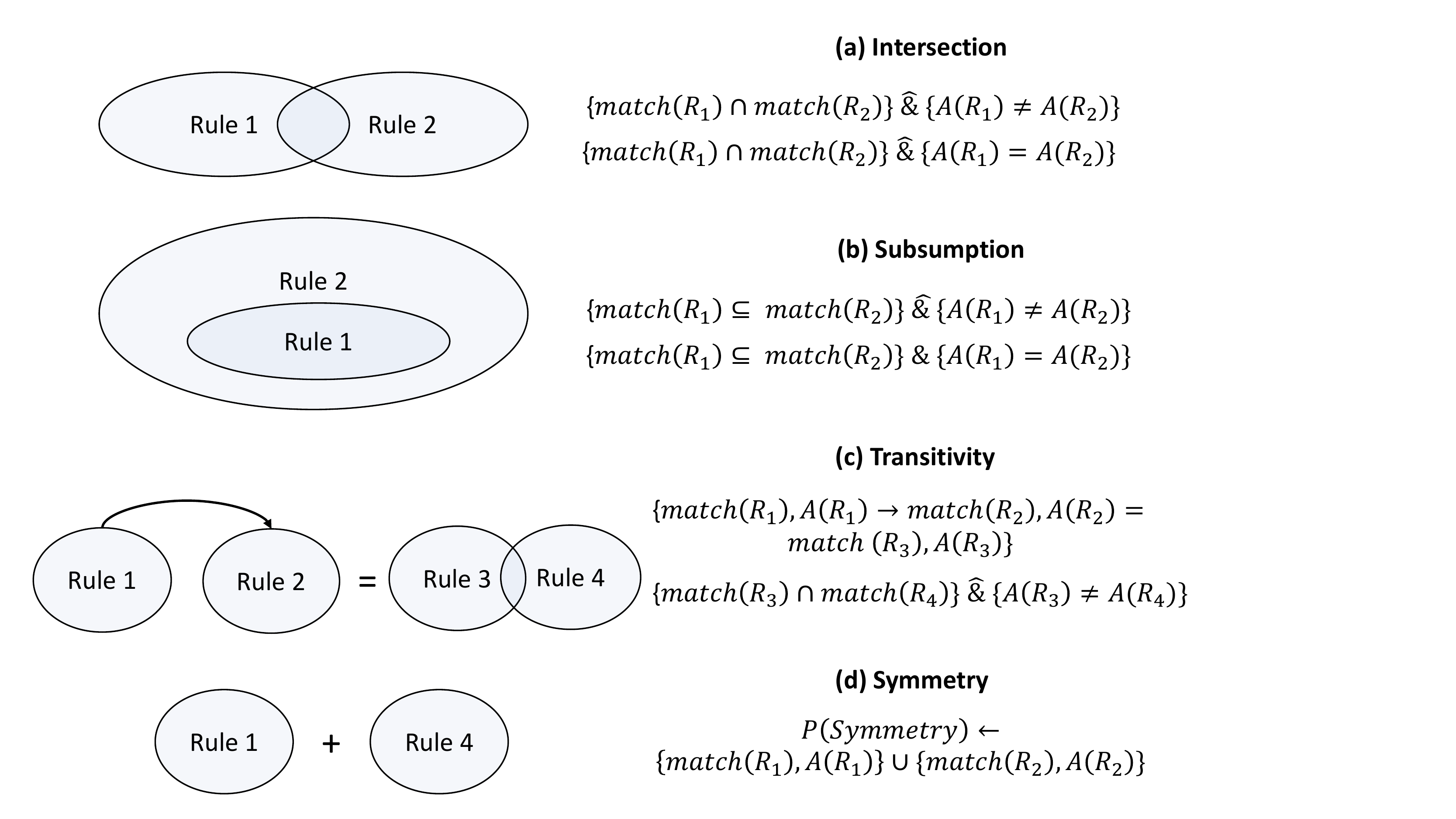}
    \caption{Security Conflict Issues in SFC}
    \label{fig:sfc06}
\end{figure}

\begin{algorithm}[ht!]
    \caption{Conflict Checking Algorithm}\label{alg02}
  \begin{small}
    \begin{algorithmic}[1]
        \Procedure{Rule Conflict Checking}{R}
        \State $R \gets$ current flow rules
        \State $R = \{match(R), A(R)\}$
        \State {C $\gets$ Conflict Set}
    
        \For{$i \in$ \{1,n\} }
        \For{$j \in$ \{1,n\} }    
        \If {$match(R_i) \cap match(R_j) \neq \emptyset$}
        \State C.add(Intersection)
        \ElsIf {$match(R_i) \subseteq match(R_j)$ OR $match(R_j) \subseteq match(R_i)$ }
        \State C.add(Subsumption)
        \ElsIf {$match(R_i) \rightarrow match(R_j) = match(R_k)$ }
        \If    {{$match(R_k) \cap match(R) \neq \emptyset$ AND $A(R_k) \cap A(R) \neq \emptyset$ }}
        \State C.add(Transitivity)
        \EndIf
        
        \ElsIf {$P(Symm) \rightarrow \{ match(R_i), A(R_i)\} \cup \{match(R_j), A(R_j)\}$ }
        \If    {{$P(Symm) \cap match(R) \neq \emptyset$}}
        \State C.add(Symmetry)
        \EndIf    
        \EndIf  
        \EndFor
        \EndFor
        \EndProcedure      
    \end{algorithmic}
    \end{small}

\end{algorithm}
\noindent  We have identified four types of conflicting scenarios that can lead to security violations,i.e., \textit{Intersection}, \textit{Subsumption}, \textit{Transitivity} and \textit{Symmetry} as shown in Figure~\ref{fig:sfc06}. The algorithm~\ref{alg02} presents the details of conflict analysis in current OpenFlow rules composed from SFs.

\begin{figure}[ht!]
	\centering
	\includegraphics[trim=0 30 0 50, width=0.45\textwidth]{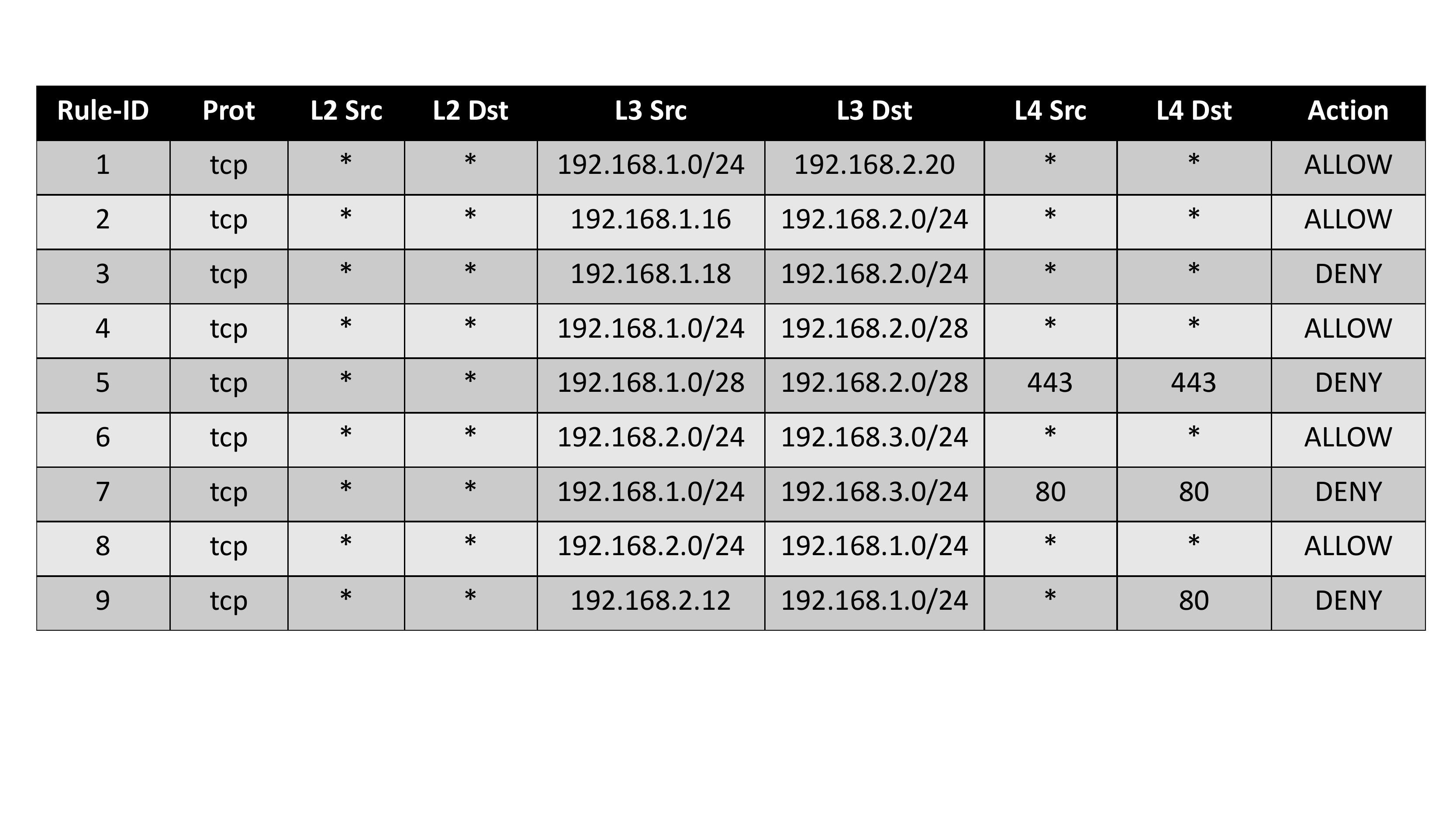}
	 \vspace{-2em}
	\caption{SFC Rule Conflict Scenarios}
	\label{tab:sfc1}
	 \vspace{-2em}
\end{figure}

\noindent The Figure~\ref{tab:sfc1} consists of OpenFlow rules which we will use to provide examples of each type of conflicting scenario in Figure~\ref{fig:sfc06}. We do not consider fields such as packet counters, timeout duration, etc. in conflict analysis. 

\noindent \textbf{Intersection} is a class of conflicts where the packet header has partial overlap across two different rules, and the actions are either same or different. For instance the rules 1 and 2 in Figure~\ref{tab:sfc1}, $l3s_2 \subset l3s_1$ and $l3d_1 \subset l3d_2$, $A(R_1) == A(R_2)$. Therefore, $\{match(R_1) \cap match(R_2)\} \neq \phi$. Similarly, for rules 1 and 3, $l3s_3 \subset l3s_1$ and $l3d_1 \subset l3d_3$, $A(R_1) \neq A(R_2)$. The match fields for rules 1 and 3 have $\{match(R_1) \cap match(R_3)\} \neq \phi$. So rules 2 and 3 have \textit{Intersection} conflict with rule 1. 

\noindent \textbf{Subsumption} refers to class of conflicts where header match of one rule is completely subsumed by another rule, and the actions are similar or dissimilar. For the rules 1 and 4, $match(R_4) \subseteq  match(R_1)$ and $A(R_4) == A(R_1)$. Another scenario of Subsumption is for the rules 1 and 5, i.e., $match(R_5) \subseteq match(R_1)$, whereas $A(R_5) \neq A(R_1)$. We classify all such conflicts in the class \textit{Subsumption}.  

\noindent \textbf{Transitivity} is a class of conflicts where the flow rule is not defined explicitly but the combination of two flow rules leads to an inferred flow rule. If the inferred flow rule can be in a state of conflict with a predefined flow rule for same header match. Consider rules 1, 6 and 7. actions for match fields $l3s_1$ = 192.168.1.0/24 and  $l3d_1$ = 192.168.2.0/24 is ALLOW. The action for rule 6, i.e. $A(R_6)$ for match fields $l3s_6$ = 192.168.1.0/24 and  $l3d_6$ = 192.168.1.0/24 is ALLOW. We can infer a new rule from these two rule, i.e., $R_1 \rightarrow R_6 = R_x$. For the rule $R_x$, the traffic between source 192.168.1.0/24 and destination 192.168.3.0/24 should be allowed by transitivity. Considering inferred rule \textit{x} and rule 7, the match fields, $match(R_7) \subseteq  match(R_x)$, however the action for rule 7, i.e. $A(R_7)$ is in conflict with action for the rule inferred - $A(R_x) \neq A(R_7)$. We classify all such scenarios into a class of transitive conflicts. 

\noindent \textbf{Symmetry} is a required property for some applications that require bi-directional connections to maintain a persistent session. For example, a stateful firewall SF requires 3-way handshake 'SYN' from source to destination, 'SYN-ACK' from destination to source and 'ACK' from source to destination. If the source address $l3s$ is 192.168.1.12 and destination address $l3d$ is 192.168.2.10, the property - symmetry  is satisfied iff $R_1$ and $R_8$ are working together, i.e., $P(Symmetry) \leftarrow R_1, R_8$. However if we check rule 9, the action for rule 9 conflicts with the action of rule 8. The $P(symmetry)$ implies $\{match(R_1) \cup match(R_8)\}$ and $A(R_1) == A(R_8)$. However, $match(R_8) \cap match(R_9) \neq \phi$ and $A(R_8) \neq A(R_9)$. Thus, rule 9 violates the symmetry property. 

\section{Implementation and Evaluation}

\subsection{Flow Composition Analysis}
\noindent We implemented Bro IDS and Linux Firewall (Netfilter) as SFs on two separate Ubuntu 16.04 VMs, between source and destination. The SFC classification policy was configured in a way that data plane traffic of the communicating machines was required to pass through the SFs. 
%

\begin{table}
    \centering
    \begin{tabular}{|c|c|c|}
        \hline
        \textbf{Time (s)} & \textbf{IDS+Netfilter Rules} & \textbf{Flow Rules} \\
        \hline
        5 & 2056 & 54 \\
        \hline
        10 & 4014 & 85 \\
        \hline
        15 & 7166 & 104 \\
        \hline
        20 & 9686 & 171 \\
        \hline
        25 & 12241 & 179 \\
        \hline
        30 & 13472 & 201 \\
        \hline    
    \end{tabular}
    \caption{IDS and Netfilter OpenFlow Rule Composition}
    \label{tab:sfc2}
    \vspace{-2em}
\end{table}
\noindent The second column in the Table~\ref{tab:sfc2} denotes number of Netfilter and IDS rules combined that are invoked in SFC $source \rightarrow Netfilter \rightarrow IDS \rightarrow destination$. It can be observed that the header fields of most of the signature-based rules present in these SFs overlap with each other and result in same OpenFlow rules. There was about 97\% reduction in number of rules ($\frac{54}{2056} \times 100 \approx$ 3\%) if the SF rules are composed into OpenFlow rules. As the time increased, from \textit{5s} to \textit{25s}, the number of SF rules invoked increase from 2056 to 13472, but the number of distinct OpenFlow rules only increase from 54 to 201. Thus, Flow composition can lead to a significant gain in terms of performance on a network with large number of SFs. 

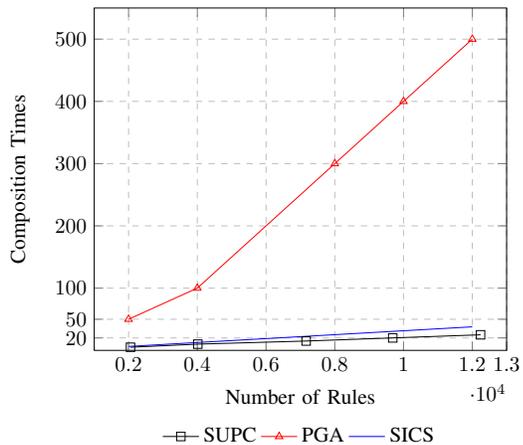
\begin{figure}[htp]
    \centering
    
    \begin{tikzpicture}[scale=0.8, transform shape]
    
    \begin{axis}[
    xlabel={Number of Rules},
    ylabel={Composition Times},
    xmin=1000, xmax=13000,
    ymin= 0, ymax= 550,
    xtick={2000, 4000, 6000, 8000, 10000, 12000, 13000},
    ytick={20, 50, 100, 200, 300, 400, 500},
    legend style={at={(0.5,-0.2)},anchor=north, legend columns=-1},
    ymajorgrids=true,
    grid=both,
    grid style=dashed,
    name=border
    ]
    
    \addplot[
    color=black,
    mark=square,
    ]
    coordinates {
        (2056, 5)(4014, 10)(7166, 15)(9686, 20)(12241, 25) 
    };
   
    \addplot[
    color=red,
    mark=triangle,
    ]
    coordinates {
        (2000, 50) (4000, 100) (8000, 300) (10000, 400) (12000, 500)
    };

    \addplot[
    color=blue,
    mark=circle,
    ]
    coordinates {
        (2000, 6.308)(4000, 12.6) (8000, 25.2) (10000, 31.5)(12000, 37.8)
    };
 
    \legend{SUPC,PGA, SICS}
    \end{axis}
    \end{tikzpicture}
    \caption{Number of Rules vs Composition Time - SUPC, PGA~\cite{prakash2015pga}, SICS~\cite{wang2016sics}}
    \label{fig:sfc09}
    \vspace{-2em}
\end{figure}
\subsubsection{Composition Time Comparitive Analysis}
\noindent We performed a comparative analysis of composition time for our algorithm~\ref{alg01} against policy composition time of PGA~\cite{prakash2015pga} and SICS framework~\cite{wang2016sics}. We use rules as a generic term to define PGA nodes, SICS rules, and OpenFlow rules, and to have a common comparison format. We observed that SUPC achieves faster composition time - 20s for 10k rules, and 25s for about 12k rules. The composition time for SICS was slightly higher than our algorithm, i.e., 31.5s for 10k rules and 37.5s for 12k rules. The composition time for PGA scales poorly with the number of rules as can be seen in the Figure~\ref{fig:sfc09}. PGA takes about 400s for the composition of 10k rules and 500s for 12k rules. The performance degradation in SICS can be attributed to encryption overhead, whereas in case of PGA, the poor scaling is because of duplication of SFs across the network. The comparison of SUPC with these frameworks shows that our flow composition algorithm will scale well with the number of SF rules.
\subsection{Flow Conflict Analysis}
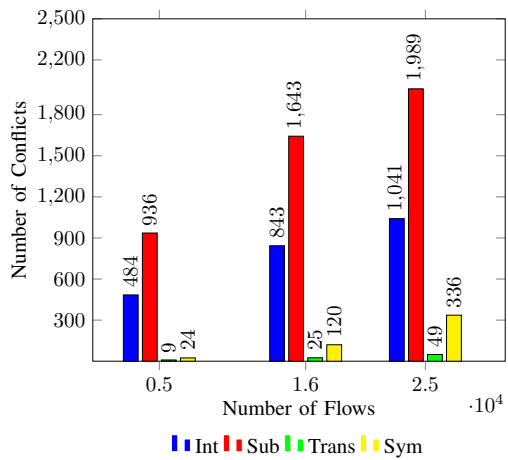
\begin{figure}[ht!]
\begin{tikzpicture}[scale=0.8, transform shape]
\begin{axis}[
ybar,
bar width=7pt,
legend style={at={(0.5,-0.2)}, anchor=north,legend columns=-1},
xmin=0,xmax=31000,
ymin=0,ymax=2500,
xlabel = Number of Flows,
ylabel= Number of Conflicts, 
enlargelimits=false, 
xtick={0,8000,16000,24000, 32000, 36000},
ytick={300, 600, 900, 1200, 1500, 1800, 2200, 2500},
xtick=data,
nodes near coords,
every node near coord/.append style={rotate=90, anchor=west},
]

\addplot [fill=blue] coordinates {(5000,484) (16000,843) (25000,1041)};

\addplot [fill=red] coordinates {(5000,936) (16000,1643) (25000,1989)}; 

\addplot [fill=green] coordinates {(5000,9) (16000,25) (25000,49)}; 

\addplot [fill=yellow] coordinates {(5000,24) (16000,120) (25000,336)}; 

\legend{Int, Sub, Trans, Sym}
\end{axis}
\end{tikzpicture}
 \vspace{-1em}
\caption{Number of Conflicts in SFC}
\label{fig:sfc07}
\end{figure}
\noindent We performed experiment to analyze the number of conflicts - \textit{Intersection (Int)}, \textit{Submsumption (Sub)}, \textit{Transitivity (Trans)} and \textit{Symmetry (Sym)} in the translated OpenFlow rules. The x-axis in the Figure~\ref{fig:sfc07} denotes the number of OpenFlow rules - 5k, 16k, and 25k. As the number of OpenFlow rules increased, we observed an increase in the number of conflicts. A remote attacker can gain access to network resources by taking advantage of these rule conflicts. The Subsumption conflicts can lead to redundant policy checks, thus leading to extra processing overhead. The Symmetry conflicts can lead to disruption of service since flow in both directions is required for stateful applications to maintain a persistent session. 


\section{Conclusion and Future Work}
\noindent The paper presents SUPC, an automated SF composition and conflict analysis framework. SUPC translates traffic and security policies of various SF into common OpenFlow format. This helps in elimination of redundant policy rules, network-wide policy enforcement using SDN controller, and conflict identification across heterogeneous SFs each having their own policy specification language. Our experimental results on the dataset of Netfilter firewall rules and Bro IDS achieved a significant reduction in matching rules $\sim 97$\% due to \textit{Flow Composition}, which leads to performance gain in SFC. We also identified four class of conflicts among the rules of various SFs which can cause security violations and service disruption. Our experiments for \textit{Flow Conflict} analysis on the dataset of an order of 1000s of rules is able to identify different possible conflicting cases. 



\section*{Acknowledgment}
This research is based upon work supported by the NRL N00173-15-G017, NSF Grants 1642031, 1528099, and 1723440, and NSFC Grants 61628201 and 61571375.

\bibliographystyle{abbrv}
\bibliography{template}

\begin{thebibliography}{10}

\bibitem{durante2017model}
L.~Durante, L.~Seno, F.~Valenza, and A.~Valenzano.
\newblock A model for the analysis of security policies in service function
  chains.
\newblock In {\em Network Softwarization (NetSoft), 2017 IEEE Conference on},
  pages 1--6. IEEE, 2017.

\bibitem{fayazbakhsh2013flowtags}
S.~K. Fayazbakhsh, V.~Sekar, M.~Yu, and J.~C. Mogul.
\newblock Flowtags: Enforcing network-wide policies in the presence of dynamic
  middlebox actions.
\newblock In {\em Proceedings of the second ACM SIGCOMM workshop on Hot topics
  in software defined networking}, pages 19--24. ACM, 2013.

\bibitem{gember2014opennf}
A.~Gember-Jacobson, R.~Viswanathan, C.~Prakash, R.~Grandl, J.~Khalid, S.~Das,
  and A.~Akella.
\newblock Opennf: Enabling innovation in network function control.
\newblock In {\em ACM SIGCOMM Computer Communication Review}, volume~44, pages
  163--174. ACM, 2014.

\bibitem{ghaznavi2016service}
M.~Ghaznavi, N.~Shahriar, R.~Ahmed, and R.~Boutaba.
\newblock Service function chaining simplified.
\newblock {\em arXiv preprint arXiv:1601.00751}, 2016.

\bibitem{joseph2008policy}
D.~A. Joseph, A.~Tavakoli, and I.~Stoica.
\newblock A policy-aware switching layer for data centers.
\newblock In {\em ACM SIGCOMM Computer Communication Review}, volume~38, pages
  51--62. ACM, 2008.

\bibitem{mckeown2008openflow}
N.~McKeown, T.~Anderson, H.~Balakrishnan, G.~Parulkar, L.~Peterson, J.~Rexford,
  S.~Shenker, and J.~Turner.
\newblock Openflow: enabling innovation in campus networks.
\newblock {\em ACM SIGCOMM Computer Communication Review}, 38(2):69--74, 2008.

\bibitem{7976378}
S.~Pisharody, J.~Natarajan, A.~Chowdhary, A.~Alshalan, and D.~Huang.
\newblock Brew: A security policy analysis framework for distributed sdn-based
  cloud environments.
\newblock {\em IEEE Transactions on Dependable and Secure Computing},
  PP(99):1--1, 2017.

\bibitem{prakash2015pga}
C.~Prakash, J.~Lee, Y.~Turner, J.-M. Kang, A.~Akella, S.~Banerjee, C.~Clark,
  Y.~Ma, P.~Sharma, and Y.~Zhang.
\newblock Pga: Using graphs to express and automatically reconcile network
  policies.
\newblock In {\em ACM SIGCOMM Computer Communication Review}, volume~45, pages
  29--42. ACM, 2015.

\bibitem{sendi2016efficient}
A.~S. Sendi, Y.~Jarraya, M.~Pourzandi, and M.~Cheriet.
\newblock Efficient provisioning of security service function chaining using
  network security defense patterns.
\newblock {\em IEEE Transactions on Services Computing}, 2016.

\bibitem{trajkovska2017sdn}
I.~Trajkovska, M.-A. Kourtis, C.~Sakkas, D.~Baudinot, J.~Silva, P.~Harsh,
  G.~Xylouris, T.~M. Bohnert, and H.~Koumaras.
\newblock Sdn-based service function chaining mechanism and service prototype
  implementation in nfv scenario.
\newblock {\em Computer Standards \& Interfaces}, 54:247--265, 2017.

\bibitem{wang2016sics}
H.~Wang, X.~Li, Y.~Zhao, Y.~Yu, H.~Yang, and C.~Qian.
\newblock Sics: Secure in-cloud service function chaining.
\newblock {\em arXiv preprint arXiv:1606.07079}, 2016.

\end{thebibliography}

\end{document}